\documentstyle[12pt]{article}
\def\A{{\cal A}}
\def\B{{\cal B}}
\def\C{{\cal C}}
\def\D{{\cal D}}
\def\1{\parallel 1>}

\def\s{\sinh}

\def\lim{\stackrel{\theta\rightarrow\pm\infty}{\sim}}

\begin{document}

\begin{titlepage}
\title{The Highest Weight property for the $SU_{q}(n)$ invariant spin chains.}

\author{H.J. de Vega \dag\\
A. Gonz\'alez--Ruiz \ddag \thanks{Work supported by the Spanish M.E.C.
under grant AP90 02620085}\\
{\it \dag L.P.T.H.E.} \\
{\it Tour 16, 1er \'etage, Universit\'e Paris VI} \\
{\it 4 Place Jussieu, 75252 Paris cedex 05, FRANCE}\\
{\it Laboratoire Associ\'e au CNRS, UA280}\\
{\it \ddag Departamento de F\'{\i}sica Te\'orica} \\
{\it Universidad Complutense, 28040 Madrid, Spain}}
\date{}
\maketitle

\begin{abstract}
The $SU_{q}(n)$ generators are obtained  as large spectral parameter limit of
the Yang-Baxter operators in the integrable $SU_{q}(n)$ invariant
vertex model. The commutation relations, including Serre
relations, are obtained as limits of the Yang-Baxter equations. The recently
found eigenvectors of the  $SU_{q}(n)$ invariant spin chains are shown to be
Highest Weight vectors of the corresponding quantum group.
\end{abstract}

\vskip-16.0cm
\rightline{{\bf F.T/U.C.M-94/xx}}
\rightline{{\bf LPTHE--PAR 94/13}}
\rightline{{\bf February 1994}}
\vskip2cm

\end{titlepage}

\begin{section}{Introduction.}
Quantum groups arose from  investigations on integrable lattice models
which provide a natural arena for its representations.
Recently the $SU_{q}(n)$ invariant spin chains \cite{ks1} have been solved
\cite{suq}. To achieve this, a non-trivial generalization of the nested Bethe
ansatz to open boundary conditions was obtained.
A Yang-Baxter (YB) algebra of the reflection type naturally appears in the
 $SU_{q}(n)$ invariant models \cite{suq,sk,lr1}. We show in this note how the
 $SU_{q}(n)$ generators for  N-sites  arise from such YB generators in
the limit when the spectral parameter tends to infinity. Then, we
prove that all Bethe states (constructed through the nested Bethe
Ansatz) are highest weight vectors. Therefore, physical states appear
as  $SU_{q}(n)$ multiplets, obtained from the Bethe states through
the action of the appropriate  generators for  N-sites.
The highest weight property is interesting in relation with the organization
of eigenstates and also in the study of the associated RSOS models
\cite{qba1}.\\
We will make some comments on the solution in order to fix the notation and
to give the necesary equations. The reader is referred to \cite{suq} for
details in the notation and some of the operators appearing in this
article.\\
The transfer matrix of the $SU_{q}(n)$ invariant chains is given
by:

\begin{equation}
t(\theta)=\sum_{a=1}^{n}K^{+}(\theta)U_{aa}(\theta).\label{t}
\end{equation}
In the previous formula $K^{+}(\theta)$ is given in \cite{suq}, and
$U_{ab}(\theta)$ is the doubled transfer matrix given by :

\begin{equation}
U_{ab}(\theta)=\sum_{c=1}^{n}T_{ac}(\theta)\tilde{T}_{cb}(\theta),\label{u}
\end{equation}
with $T,\;\tilde{T}$ Yang-Baxter operators. The doubled transfer matrix can
be written as an operator matrix with elements $U_{11}=\A,\;U_{1j}=\B_{j},\;
U_{j1}=\C_{j},\;U_{jk}=\D_{jk};\;j,k=2,\ldots,n$. In the solution of the
eigenvalue problem for the transfer matrix (\ref{t}) the following operators
turn out to be useful and will appear in section 3:

\begin{eqnarray}
\hat{\D}_{bd}(\theta)&=&\frac{1}{\s 2\theta}[e^{2\theta}\s(2\theta+\gamma)
\D_{bd}(\theta)-\s\gamma\delta_{bd}\A(\theta)]\nonumber\\
\hat{\B}_{c}(\theta)&=&\frac{\s(2\theta+\gamma)}{\s 2\theta}\;\B_{c}(\theta).
\nonumber
\end{eqnarray}

As Bethe ansatz for the eigenstates of the transfer matrix (\ref{t}) we use:

\begin{equation}
\Psi \equiv\sum_{2\leq i_{j}\leq
n}X^{i_{1}\ldots i_{p_{1}}}\hat{\B}_{i_{1}}(\mu_{1})\ldots
\hat{\B}_{i_{p_{1}}}(\mu_{p_{1}})\1\nonumber\\
=\hat{\B}(\mu_{1})\otimes\ldots\otimes\hat{\B}(\mu_{p_{1}})\1 X.
\label{bare}
\end{equation}

Parallel to the case of periodic boundary conditions the
system is solved by giving a recursion relation between the original
eigenvalue and the one of a  reduced problem with one less state per link.
This relation is given by:

\begin{eqnarray}
&&\Lambda^{(k)}(\theta,\tilde{\mu}^{(k-1)})=\prod_{j=1}^{p_{k}}
\frac{\s[\theta+\mu_{j}^{(k)}+(k-1)\gamma]\s(\theta-\mu_{j}^{(k)}-\gamma)}
{\s(\theta+\mu_{j}^{(k)}+k\gamma)\s(\theta-\mu_{j}^{(k)})}\nonumber\\
&&+\frac{\s[2\theta+(k-1)\gamma]}{\s[2\theta+(k+1)\gamma]}\nonumber\\
&&\prod_{j=1}^{p_{k-1}}
\frac{\s[\theta+\mu_{j}^{(k-1)}+(k-1)\gamma]\s(\theta-\mu_{j}^{(k-1)})}
{\s(\theta+\mu_{j}^{(k-1)}+k\gamma)\s(\theta-\mu_{j}^{(k-1)}+\gamma)}
\nonumber\\
&&\prod_{j=1}^{p_{k}}
\frac{\s[\theta+\mu_{j}^{(k)}+(k+1)\gamma]\s(\theta-\mu_{j}^{(k)}+\gamma)}
{\s(\theta+\mu_{j}^{(k)}+k\gamma)\s(\theta-\mu_{j}^{(k)})}
\Lambda^{(k+1)}(\theta,\mu^{(k)}),
\label{rec}\\
&&1\leq k\leq n-1\;\;,\mu_{j}^{(0)}=0
\;\;,\Lambda^{(n)}(\theta,\mu^{(n-1)})=1,\;\;p_{0}=N.\nonumber
\end{eqnarray}

The roots $\mu^{(k)}_{i}$ have to obey the Bethe ansatz
equations:

\begin{eqnarray}
&&\Lambda^{(k+1)}(\mu_{i}^{(k)},\tilde{\mu}^{(k)})=\nonumber\\
&&\prod_{j=1}^{p_{k-1}}
\frac{\s(\mu_{i}^{(k)}+\mu_{j}^{(k-1)}+k\gamma)
\s(\mu_{i}^{(k)}-\mu_{j}^{(k-1)}+\gamma)}
{\s[\mu_{i}^{(k)}+\mu_{j}^{(k-1)}+(k-1)\gamma]
\s(\mu_{i}^{(k)}-\mu_{j}^{(k-1)})}\nonumber\\
&&\prod^{p_{k}}_{j\neq i}\frac{\s[\mu_{i}^{(k)}+\mu_{j}^{(k)}+(k-1)\gamma]
\s(\mu_{i}^{(k)}-\mu_{j}^{(k)}-\gamma)}
{\s[\mu_{i}^{(k)}+\mu_{j}^{(k)}+(k+1)\gamma]
\s(\mu_{i}^{(k)}-\mu_{j}^{(k)}+\gamma)},
\label{polk}
\end{eqnarray}
where we have set the inhomogeneities at the first level equal to zero
 for simplicity (notice also the change $\theta\rightarrow\theta+\gamma/2$
with respect to reference \cite{suq}).\\
Using the recursion formula (\ref{rec}) the eigenvalue problem is solved.

\end{section}

\begin{section}{The $SU_{q}(n)$ generators and its relations.}
In this section we show how in certain limits of the spectral parameter
$\theta$ the Yang--Baxter algebra leads to the quantum group $SU_{q}(n)$.\\
First we obtain the generators of the quantum algebra as the leading terms of
the transfer matrices $T,\tilde{T}$ in the limits $\theta\rightarrow\pm\infty$.
We find ($q=e^{\gamma}$):

\begin{eqnarray}
&&T_{ab}(\infty):=T^{+}=\nonumber\\
&&=\left\{\begin{array}{ll}
a>b;&0,\\
a=b;&q^{-L} q^{W_{a}},\\
a<b;&\left\{\begin{array}{ll}
b=a+1;&q^{-L}(q-q^{-1})q^{-1/2}J^{-}_{a}q^{W_{a+1}/2}q^{W_{a}/2},\\
b=a+j,\;j>1;&T^{j}_{-},
\end {array}\right.
\end {array}\right.
\label{gem}\\
&&T_{ab}(-\infty):=T^{-}=\nonumber\\
&&=\left\{\begin{array}{ll}
a>b;&\left\{\begin{array}{ll}
a=b+1;&-q^{L}(q-q^{-1})q^{1/2}J^{+}_{a-1}q^{-W_{a-1}/2}q^{-W_{a}/2},\\
a=b+j,\;j>1;&T^{j}_{+},
\end {array}\right.\\
a=b;&q^{L} q^{-W_{a}},\\
a<b;&0,
\end {array}\right.
\label{gep}\\
&&\tilde{T}_{ab}(\infty)=\nonumber\\
&&=\left\{\begin{array}{ll}
a>b;&\left\{\begin{array}{ll}
a=b+1;&q^{-L}(q-q^{-1})q^{-1/2}q^{W_{a}/2}q^{W_{a-1}/2}J^{+}_{a-1},\\
a=b+j,\;j>1;&\tilde{T}^{j}_{+},
\end {array}\right.\\
a=b;&q^{-L} q^{W_{a}},\\
a<b;&0,
\end {array}\right.
\label{getp}\\
&&\tilde{T}_{ab}(-\infty)=\nonumber\\
&&=\left\{\begin{array}{ll}
a>b;&0,\\
a=b;&q^{L} q^{-W_{a}},\\
a<b;&\left\{\begin{array}{ll}
b=a+1;&-q^{L}(q-q^{-1})q^{1/2}q^{-W_{a}/2}q^{-W_{a+1}/2}J^{-}_{a},\\
b=a+j,\;j>1;&\tilde{T}^{j}_{-}.
\end {array}\right.
\end {array}\right.
\label{getm}
\end{eqnarray}

In the previous formulas the following operators have been introduced:

\begin{eqnarray}
&&q^{\pm W_{a}}=q^{\pm E_{aa}}\otimes q^{\pm E_{aa}}\otimes\ldots
\otimes q^{\pm E_{aa}}\nonumber\\
&&J^{+}_{a}=\sum_{i=1}^{L}q^{- h_{a}/2}\otimes\ldots q^{- h_{a}/2}\otimes
E_{aa+1}^{i_{th}}\otimes q^{h_{a}/2}\otimes\ldots\otimes q^{h_{a}/2}
\label{qg}\\
&&J^{-}_{a}=\sum_{i=1}^{L}q^{- h_{a}/2}\otimes\ldots q^{- h_{a}/2}\otimes
E_{a+1a}^{i_{th}}\otimes q^{h_{a}/2}\otimes\ldots\otimes q^{h_{a}/2},
\nonumber
\end{eqnarray}
where $[E_{ab}]_{ij}=\delta_{ia}\delta_{jb}$ and
$h_{a}=E_{aa}-E_{a+1a+1}$ are the $su(n)$ generators in the fundamental
representation. The operators $T^{j}_{\pm}$, corresponding to the limits of
the $T_{ab}$ operator, are polynomials of order $j$ on the generators
$J^{\pm}_{l},\;l=b,b\pm 1, \ldots,a\mp 1$. Formula (\ref{qg}) gives the
coproduct of the the quantum group generators to the power $L-1$ \cite{jim},
we will see that this gives in fact a representation of the quantum group on
the lattice of $L$ sites.\\  The operators $T_{ab}(\theta)$ obey the
Yang-Baxter relation:

\begin{eqnarray}
M^{ab}_{cd}=R(\theta-\theta^{\prime})^{ab}_{ef}T(\theta)_{ec}
T(\theta^{\prime})_{fd}=
T(\theta^{\prime})_{ae}T(\theta)_{bf}R(\theta-\theta^{\prime})^{ef}_{cd}
=N^{ab}_{cd}
\label{yb}.
\end{eqnarray}

By taking the limit $\theta\rightarrow-\infty,
\theta^{\prime}\rightarrow\infty$, in the
previous equations we get the spectral parameter independent commutation
relations:

\begin{eqnarray}
R_{-}{ }^{ab}_{ef}T^{-}_{ec}
T^{+}_{fd}=
T^{+}_{ae}T^{-}_{bf}R_{-}{ }^{ef}_{cd}
\nonumber,
\end{eqnarray}
where $R_{-}={\rm lim}_{\theta\rightarrow-\infty}R(\theta)$.
If we use the equations $M^{ab+1}_{ba+1}=N^{ab+1}_{ba+1}$
the following
commutation relations are obtained:

\begin{eqnarray}
&&[J^{+}_{i},J^{-}_{j}]=\delta_{ij}\frac{q^{H_{i}}-q^{-H_{i}}}{q-q^{-1}},
\label{qg1}\\
&&q^{H_{i}}=q^{h_{i}}\otimes\ldots\otimes q^{h_{i}}
\nonumber.
\end{eqnarray}

Using $M^{ab+1}_{ba}=N^{ab+1}_{ba}$ and $M^{ab}_{b+1\;a}=
N^{ab}_{b+1\;a}$ the result
is:

\begin{eqnarray}
q^{H_{i}}J^{\pm}_{j}q^{-H_{i}}=q^{\pm a_{ij}}J^{\pm}_{j}
\label{qg2},
\end{eqnarray}

where $(a_{ij})_{1\leq i,j\leq n-1}$ denotes the Cartan matrix of type
$A_{n-1}$, i.e., $a_{ii}=2,\;a_{ij}=-1\;(i=j\pm1),\;0$ (otherwise).\\
In the limit $\theta\rightarrow-\infty,\;\theta^{\prime}\rightarrow-\infty$,
the the spectral parameter independent Yang-Baxter relation is:

\begin{eqnarray}
R_{-}{ }^{ab}_{ef}T^{-}_{ec}
T^{-}_{fd}=
T^{-}_{ae}T^{-}_{bf}R_{-}{ }^{ef}_{cd}
\nonumber.
\end{eqnarray}

Using the equalities $M^{a+1b+1}_{ba}=N^{a+1b+1}_{ba},\;b\neq a,a\pm1$ the
result is:

\begin{eqnarray}
J^{+}_{a}J^{+}_{b}=J^{+}_{b}J^{+}_{a},\; \mid a-b\mid\geq 2
\label{qg3}.
\end{eqnarray}

When $b=a+1$ in the previous equality:

\begin{eqnarray}
T(-\infty)_{a+2a}=(q-q^{-1})q^{L}q^{3/2}[J^{+}_{a+1}J^{+}_{a}-q^{-1}
J^{+}_{a}J^{+}_{a+1}]q^{-W_{a}/2}q^{-W_{a+2}/2}\label{lin},
\end{eqnarray}
and using $M^{a+1a+2}_{aa}=N^{a+1a+2}_{aa}$ we obtain the Serre relation:

\begin{eqnarray}
(J^{+}_{a})^{2}J^{+}_{a+1}-(q+q^{-1})J_{a}^{+}J^{+}_{a+1}J_{a}^{+}+
J^{+}_{a+1}(J^{+}_{a})^{2}=0,\;1\leq a,a+1\leq n-1\label{qg4}.
\end{eqnarray}

Making $a\rightarrow a-1$ in (\ref{lin}) and using
$M^{a+1a+1}_{a-1a}=N^{a+1a+1}_{a-1a}$ a second Serre relation is obtained:

\begin{eqnarray}
(J^{+}_{a})^{2}J^{+}_{a-1}-(q+q^{-1})J_{a}^{+}J^{+}_{a-1}J_{a}^{+}+
J^{+}_{a-1}(J^{+}_{a})^{2}=0,\;1\leq a,a-1\leq n-1\label{qg5}.
\end{eqnarray}

Proceeding in a parallel way with the limit
$\theta\rightarrow\infty,\;\theta^{\prime}\rightarrow\infty$ of relation
(\ref{yb}) given by:

\begin{eqnarray}
R_{+}{ }^{ab}_{ef}T^{+}_{ec}
T^{+}_{fd}=
T^{+}_{ae}T^{+}_{bf}R_{+}{ }^{ef}_{cd}
\nonumber,
\end{eqnarray}
where $R_{+}={\rm lim}_{\theta\rightarrow \infty}R(\theta)$,
the rest of relations are obtained:

\begin{eqnarray}
&&q^{H_{a}}q^{H_{b}}=q^{H_{b}}q^{H_{a}}\label{qg6}\\
&&J^{-}_{a}J^{-}_{b}=J^{-}_{b}J^{-}_{a},\; \mid a-b\mid\geq 2\label{qg7}\\
&&(J^{-}_{a})^{2}J^{-}_{a\pm1}-(q+q^{-1})J_{a}^{-}J^{-}_{a\pm1}J_{a}^{-}+
J^{-}_{a\pm1}(J^{-}_{a})^{2}=0,\;1\leq a,a\pm1\leq n-1\label{qg8}.
\end{eqnarray}

Equations (\ref{qg1}-\ref{qg8}) give the $SU_{q}(n)$ quantum group
commutation relations \cite{jim}.\\
We pass at this point to the study  of the large spectral parameter limits
of the doubled monodromy matrix. Using equations (\ref{gem}-\ref{getm}) and
the definition of the doubled monodromy matrix we have:

\begin{eqnarray}
U_{ab}(\infty)=\sum_{l\geq {\sl max}(a,b)}T_{al}(\infty)
\tilde{T}_{lb}(\infty)
\nonumber\\
U_{ab}(-\infty)=\sum_{l\leq{\sl min}(a,b)}T_{al}(-\infty)
\tilde{T}_{lb}(-\infty).\label{lim}
\end{eqnarray}

{F}rom these formulas we see that not all the  quantum group generators can
be obtained cleanly from the limits of the doubled monodromy matrix. We will
have in general that these limits are formed by polinomials in the
generators.
Nevertheless for some special cases is possible to obtain cleanly the quantum
group generators, these are:

\begin{eqnarray}
&&U_{nn}(\infty)=q^{-2L}q^{2W_{n}}\nonumber\\
&&U_{n-1n}(\infty)=q^{-2L}(q-q^{-1})q^{-1/2}J_{n-1}^{-}q^{3W_{n}/2}q^{W_{n-1}/2}
\nonumber\\
&&U_{nn-1}(\infty)=q^{-2L}(q-q^{-1})q^{-1/2}q^{3W_{n}/2}q^{W_{n-1}/2}J_{n-1}^{+}
\nonumber\\
&&U_{11}(-\infty)=q^{2L}q^{-2W_{1}}\nonumber\\
&&U_{12}(-\infty)=-q^{2L}(q-q^{-1})q^{1/2}q^{-3W_{1}/2}q^{-W_{2}/2}J_{1}^{-}
\nonumber\\
&&U_{21}(-\infty)=-q^{2L}(q-q^{-1})q^{1/2}J_{1}^{+}q^{-3W_{1}/2}q^{-W_{2}/2}
\label{uinf}
\end{eqnarray}

\end{section}

\begin{section}{The Highest Weight property.}
In this section we prove the highest weight property for the Bethe
eigenstates. This property has been  shown to hold in the case of open
spin chains for the $SU_{q}(2)$ invariant case, for the $SO(4)$ Hubbard
model and   for the
$spl_{q}(2,1)$ invariant t-J model \cite{qba1,lr2,eks,tj}. This is the first
proof
for an algebra of arbitrary rank.\\
We need to obtain the commutation relation between the infinite spectral
parameter limits of the doubled monodromy matrix and the operators
$\hat{\B}$. For that we use the ``reflection'' relation \cite{suq,sk}:

\begin{eqnarray}
\begin{array}{l}
M^{ab}_{cd}\equiv R(\theta-\theta^\prime)^{ab}_{ef}\;U_{eg}(\theta)R(\theta
 +\theta^\prime)^{gf}_{hd}\;U_{hc}(\theta^\prime)=\\
N^{ab}_{cd}\equiv U_{ae}(\theta^\prime)\;R(\theta +\theta^\prime)^{eb}_{fg}
\;U_{fh}(\theta)\;R(\theta - \theta^\prime)^{hg}_{cd}
\end{array}\nonumber,
\end{eqnarray}
which in some special values of $a,b,c,d$ and limits will give the necessary
relations. It is necessary to prove that
$J_{a}^{+}\Psi=0,\;a=1,\ldots,n-1$.\\
We begin proving the case $J_{1}^{+}\Psi=0$.\\
Take the equation $M^{21}_{1l}=N^{21}_{1l}$ in the limit
$\theta^\prime\rightarrow-\infty$ to obtain:

\begin{eqnarray}
J_{1}^{+}q^{-3W_{1}/2}q^{-W_{2}/2}\hat{B}_{l}(\theta)=
R_{-}{
}^{2j}_{l2}\hat{B}_{j}(\theta)J_{1}^{+}q^{-3W_{1}/2}q^{-W_{2}/2}\nonumber\\
+q^{3/2}[\delta_{l2}A(\theta)-e^{-2\theta}\hat{\D}_{2l}(\theta)]q^{-2W_{1}}.
\label{com1}
\end{eqnarray}
Using
$M^{11}_{1l}=N^{11}_{1l}$ in the $\theta^\prime\rightarrow-\infty$ limit:

\begin{eqnarray}
q^{-2W_{1}}\hat{B}_{l}(\theta)=q^{2}\hat{B}_{l}(\theta)q^{-2W_{1}}.
\label{com1b}
\end{eqnarray}

As $J_{1}^{+}\1=0$, with the help of
 commutation relations (\ref{com1},
\ref{com1b})
we find:

\begin{eqnarray}
&&J_{1}^{+}q^{-3W_{1}/2}q^{-W_{2}/2}\Psi=q^{2(P_{1}-L-1/4)}\sum_{k=1}^{p_{1}}
\delta_{j_{k}2}
\left[\prod_{j\neq
k}^{p_{1}}\frac{\s(\mu^{(1)}_{k}+\mu^{(1)}_{j})
\s(\mu^{(1)}_{k}-\mu^{(1)}_{j}-\gamma)}
{\s(\mu^{(1)}_{k}+\mu^{(1)}_{j}+\gamma)\s(\mu^{(1)}_{k}-\mu^{(1)}_{j})}
\right.\nonumber\\
&&-\prod_{j\neq
k}^{p_{1}}\frac{\s(\mu^{(1)}_{k}+\mu^{(1)}_{j}+2\gamma)
\s(\mu^{(1)}_{k}-\mu^{(1)}_{j}+\gamma)}
{\s(\mu^{(1)}_{k}+\mu^{(1)}_{j}+\gamma)
\s(\mu^{(1)}_{k}-\mu^{(1)}_{j})}\nonumber\\
&&\left.\prod_{i=1}^{L}\left[\frac{\s\mu^{(1)}_{k}}
{\s(\mu^{(1)}_{k}+\gamma)}\right]^{2}
\Lambda^{(2)}(\mu^{(1)}_{k};\tilde{\mu}^{(1)})\right]\nonumber\\
&&\hat{\B}_{j_{k+1}}(\mu^{(1)}_{k+1}\ldots\hat{\B}_{j_{k-1}}(\mu^{(1)}_{k-1}
\1 M^{(j)}_{(i)}X^{(i)}
=0,\label{hw1}
\end{eqnarray}
where $M^{(j)}_{(i)}$ takes count of the reordering of the $\tilde{\B}$
operators \cite{suq,hr}. The last equality in the previous equation holds by
virtue of the first level Bethe ansatz equations. Using that
$J_{1}^{+}q^{-3W_{1}/2}q^{-W_{2}/2}=q q^{-3W_{1}/2}q^{-W_{2}/2}J_{1}^{+}$
and that $q^{W_{i}}$ are invertible operators, this prove that
$J_{1}^{+}\Psi=0$.\\
For the rest of the generators things are not so easy. First we will prove
that:

\begin{equation}
U_{bd}(\infty)\Psi=0,\;\;b>d,\;\;b,d>1
\end{equation}
Using the equations $M_{cd}^{1b}=N^{1b}_{cd}$ and taking the limit
$\theta\rightarrow\infty$ (see appendix A of \cite{suq} for details):

\begin{eqnarray}
U_{bd}(\infty)\Psi=q^{2 p_{1}}\hat{\B}(\mu^{(1)}_{1})\otimes
\ldots\otimes\hat{\B}(\mu^{(1)}_{p_{1}})\1 U^{(2)}_{bd}(\infty)X.
\end{eqnarray}
In the previous formula $U{ }^{(2)}_{bd}(\infty)$ is the
$\theta\rightarrow\infty$ limit of the $U_{bd}(\theta+\gamma/2)$ operator of
a problem with local weights
$R^{(2)}{ }^{ij}_{km}(\theta+\gamma/2),\;i,j,k,m=2,\ldots,n$. It can be seen
also as a $U_{b-1d-1}(\theta+\gamma/2)$ operator of an $SU_{q}(n-1)$ chain
with local weights $R^{ij}_{km}(\theta+\gamma/2),\;i,j,k,m=1,\ldots,n-1$. We
can follow this process $l$ times up to the moment when $d-l=1$:

\begin{eqnarray}
U_{bd}(\infty)\Psi=q^{2(p_{1}+p_{2}+\ldots
+p_{d-1})}\hat{\B}(\mu^{(1)}_{1})\otimes
\ldots\otimes\hat{\B}^{(d-1)}(\mu^{(d-1)}_{j_{p_{d-1}}})\parallel 1^{(l)}>
U^{(d)}_{bd}(\infty)X^{(d-1)}.\nonumber
\end{eqnarray}

Now $U^{(d)}_{bd}(\infty)$ can be seen as
$U_{b-l1}(\theta+l\gamma/2)=C_{b-l}(\theta+l\gamma/2)$ operator for an
$SU_{q}(n-l)$ chain with local weights
$R^{ij}_{km}(\theta+l\gamma/2),\;i,j,k,m=1,\ldots,n-l$. We arrive with
this process always to an operator of the type $C_{b-l}(\infty)$ at
level $l+1$ acting on $X^{(l)}$. We need to now the commutation relations
between operators $C_{b}(\infty)$ and  $\hat{\B}_{c}(\theta)$ to evaluate
this action.
The commutation relations are obtained in the $\theta\rightarrow\infty$ limit
of the relation $M^{1b}_{c1}=N^{1b}_{c1}$, the result is:

\begin{eqnarray}
C_{b}(\infty)\hat{\B}_{c}(\theta)=R_{+}{ }^{bi}_{cj}\hat{\B}_{i}(\theta)
C_{j}(\infty)\nonumber\\
+q(q-q^{-1})D_{bg}(\infty)[\delta_{gc}\A(\theta)-e^{-2\theta}
\hat{\D}_{gc}(\theta)],
\end{eqnarray}
where a summation in $g$ is understood and we have used
$[D_{bg}(\infty),\A(\theta)]=0$. Using this formula at an arbitrary level:

\begin{eqnarray}
&&U^{(d)}_{bd}(\infty)X^{(d-1)}=
q(q-q^{-1})\sum_{k=1}^{p_{d}}D^{(d)}_{bj_{k}}(\infty)\nonumber\\
&&\left[\prod_{j\neq
k}^{p_{d}}\frac{\s[\mu^{(d)}_{k}+\mu^{(d)}_{j}+(d-1)\gamma]
\s(\mu^{(d)}_{k}-\mu^{(d)}_{j}-\gamma)}
{\s(\mu^{(d)}_{k}+\mu^{(d)}_{j}+d\gamma)\s(\mu^{(d)}_{k}-\mu^{(d)}_{j})}-
\right.\nonumber\\
&&\prod_{j\neq
k}^{p_{d}}\frac{\s[\mu^{(d)}_{k}+\mu^{(d)}_{j}+(d+1)\gamma]
\s(\mu^{(d)}_{k}-\mu^{(d)}_{j}+\gamma)}
{\s(\mu^{(d)}_{k}+\mu^{(d)}_{j}+d\gamma)
\s(\mu^{(d)}_{k}-\mu^{(d)}_{j})}\nonumber\\
&&\left.\prod_{i=1}^{p_{d-1}}
\frac{\s[\mu^{(d)}_{k}+\mu^{(d-1)}_{i}+(d-1)\gamma]
\s(\mu^{(d)}_{k}-\mu^{(d-1)}_{i})}
{\s(\mu^{(d)}_{k}+\mu^{(d-1)}_{i}+d\gamma)
\s(\mu^{(d)}_{k}-\mu^{(d-1)}_{i}+\gamma)}
\Lambda^{(d+1)}(\mu^{(d)}_{k};\tilde{\mu}^{(d)})\right]\nonumber\\
&&\hat{\B}^{(d)}_{j_{k+1}}(\mu^{(d)}_{k+1}\ldots
\hat{\B}^{(d)}_{j_{k-1}}(\mu^{(d)}_{k-1})
\parallel 1^{(d)}> M^{(j)}_{(i)}X^{(d)}{ }^{(i)}\nonumber\\
&&=0,\label{hwd}
\end{eqnarray}
where the last equality holds by virtue of the Bethe ansatz equations at level
 $d$. In conclusion this shows the desired result
$U_{bd}(\infty)\Psi=0, \;b>d,\;b,d>1$. This shows directly with
(\ref{uinf}) that $J^{+}_{n-1}\Psi=0$.
Using this result
$U_{nn-2}(\infty)\Psi=T_{nn}(\infty)\tilde{T}_{nn-2}(\infty)\Psi=0$, as
$T_{nn}(\infty)$ is an invertible operator then
$\tilde{T}_{nn-2}(\infty)\Psi=0$. Now using
$U_{n-1n-2}(\infty)\Psi=[T_{n-1n-1}(\infty)\tilde{T}_{n-1n-2}(\infty)
+T_{n-1n}(\infty)\tilde{T}_{nn-2}(\infty)]\Psi=0$, which implies
$\tilde{T}_{n-1n-2}(\infty)\Psi=0$, and the relations (\ref{getp}) we
have $J^{+}_{n-2}\Psi=0$. In this way it is possible to prove the highest
weight property for the rest of the generators.\\
This ends the proof of the highest weight
property $J^{+}_{l}\Psi=0,\;l=1,\ldots,n-1$

We have seen that the highest weight property of the Bethe states is
maintained in the open quantum group invariant case for an algebra of
arbitrary rank. The quantum group generators and relations are obtained in
the usual way. This proof is interesting in the study of the associated RSOS
models and is currently under investigation. It would be interesting to
generalize this property to the cases where an elliptic symmetry of the
model is present as in \cite{ff}.

\end{section}

\end{document}